\documentclass[twocolumn]{aastex631}

\shorttitle{unWISE Time-Domain Catalog}
\shortauthors{Meisner et al.}

\begin{document}

\title{unTimely: a Full-sky, Time-Domain unWISE Catalog}

\correspondingauthor{Aaron M. Meisner}
\email{aaron.meisner@noirlab.edu}

\author[0000-0002-1125-7384]{Aaron M. Meisner}
\affiliation{NSF's National Optical-Infrared Astronomy Research Laboratory, 950 N. Cherry Ave., Tucson, AZ 85719, USA}

\author[0000-0001-7896-5791]{Dan Caselden}
\affiliation{Department of Astrophysics, American Museum of Natural History, Central Park West at 79th Street, NY 10024, USA}

\author[0000-0002-3569-7421]{Edward F. Schlafly}
\affiliation{Space Telescope Science Institute, 3700 San Martin Drive, Baltimore, MD 21218, USA}

\author[0000-0001-8662-1622]{Frank Kiwy}
\affiliation{Backyard Worlds: Planet 9}

\begin{abstract}

We present the unTimely Catalog, a deep time-domain catalog of detections based on Wide-field Infrared Survey Explorer (WISE) and NEOWISE observations spanning the 2010 through 2020 time period. Detections are extracted from `time-resolved unWISE coadds', which stack together each biannual sky pass of WISE imaging to create a set of $\sim$16 all-sky maps (per band), each much deeper and cleaner than individual WISE exposures. unTimely incorporates the W1 (3.4~$\mu$m) and W2 (4.6~$\mu$m) channels, meaning that our data set effectively consists of $\sim$32 full-sky unWISE catalogs. We run the \verb|crowdsource| crowded-field point source photometry pipeline \citep{decaps} on each epochal coadd independently, with low detection thresholds: S/N = 4.0 (2.5) in W1 (W2). In total, we tabulate and publicly release 23.5 billion (19.9 billion) detections at W1 (W2). unTimely is $\sim$1.3 mag deeper than the WISE/NEOWISE Single Exposure Source Tables near the ecliptic, with further enhanced depth toward higher ecliptic latitudes. The unTimely Catalog is primarily designed to enable novel searches for faint, fast-moving objects, such as Y dwarfs and/or late-type (T/Y) subdwarfs in the Milky Way's thick disk or halo. unTimely will also facilitate other time-domain science applications, such as all-sky studies of quasar variability at mid-infrared wavelengths over a decade-long time baseline.

\end{abstract}


\section{Introduction} \label{sec:intro}

The Wide-field Infrared Survey Explorer \citep[WISE;][]{wright10} provides an unprecedented time-domain view of the mid-infrared sky, offering a decade plus time baseline at W1 = 3.4~$\mu$m and W2 = 4.6~$\mu$m thanks to the NEOWISE mission extension \citep{neowise,neowiser}. As of this writing, WISE has completed more than 19 full-sky mappings over the course of $> 12.5$ years, with each sky location observed in $\gtrsim$12 single exposures (per band) spanning a $\sim$1 day time interval during each biannual sky pass. The public archive of WISE/NEOWISE single-exposure source extractions contains a remarkable $> 165$ billion detections \citep{cutri15}. However, the WISE/NEOWISE Single Exposure (L1b) Source Tables are limited by the relatively shallow per-exposure depth, and inevitably contain many contaminants such as satellite streaks and cosmic rays. By coadding WISE observations on a per sky pass basis, it is possible to reach $\sim$1.3 magnitudes ($2.5\times\textrm{log}_{10}\sqrt{12}$) or more deeper while also nulling out short-lived artifacts, at the expense of losing temporal information on timescales $\lesssim $1 day.

Although upcoming projects such as NEO Surveyor \citep{neo_surveyor,davy_white_paper,ross_neo_surveyor} and Rubin Observatory's Legacy Survey of Space and Time \citep[LSST;][]{lsst} promise exciting new hauls of wide-area time series data, no current or planned mission will surpass WISE in terms of all-sky imaging at mid-infrared wavelengths. It is therefore critical to maximally mine the WISE archive, especially given WISE's complementarity with JWST \citep{jwst} and future surveys including SPHEREx \citep{spherex}, Euclid \citep{euclid}, Rubin/LSST, NEO surveyor, and Roman \citep{roman}.

Our ``unWISE'' archival data analysis effort (\url{http://unwise.me}) seeks to fully realize the potential of WISE/NEOWISE imaging for extragalactic and Galactic science \citep{lang14,lang14b,fulldepth_neo1, fulldepth_neo2, tr_neo2,fulldepth_neo3,tr_neo3,neo4_coadds,unwise_catalog,dey_overview,fulldepth_neo5,fulldepth_neo6,fulldepth_neo7}. In particular, our `time-resolved unWISE coadds' \citep{tr_neo2}, which stack exposures on a per sky pass basis, provide a novel, deep, and clean time-domain view of the mid-infrared sky. But until now there has been no full-sky catalog based on these time-resolved unWISE coadds. Here we remedy this situation by creating and publicly releasing a full-sky catalog built from the unWISE time-resolved coadds spanning 2010 through 2020, which we refer to as the ``unTimely Catalog''. Typically, a given sky location has a set of 16 epochal unTimely catalogs per-band, each much deeper and cleaner than the WISE/NEOWISE Single Exposure Source Tables. We therefore expect the unTimely Catalog to enable studies/discoveries of cool moving objects \citep[e.g., nearby brown dwarfs;][]{,pinfield_methodology,backyard_worlds} and flux variables \citep[e.g., quasars and young stellar objects;][]{kozlowski,ysos} to depths $\gtrsim 1.3$ mag fainter than would be possible with individual WISE exposures.

In $\S$\ref{sec:coadds} we describe the set of time-resolved unWISE coadds that forms the input for our unTimely Catalog source extraction. In $\S$\ref{sec:crowdsource} we discuss our deployment of the \verb|crowdsource| source detection/photometry software \citep{decaps, crowdsource_ascl} on the time-resolved unWISE coadd data set. In $\S$\ref{sec:completeness} we evaluate the completeness/reliability of our unTimely catalogs. In $\S$\ref{sec:zp} we investigate unTimely's photometric zeropoint consistency over time and bright end scatter. In $\S$\ref{sec:sci_apps} we highlight example unTimely Catalog science use cases. In $\S$\ref{sec:dr} we describe our unTimely Catalog data release. In $\S$\ref{sec:caveats} we list cautionary notes regarding the use of our catalogs and outline some potential improvements/features that may be implemented for future data releases. We conclude in $\S$\ref{sec:conclusion}.

\section{Input Time-Resolved unWISE Coadds} \label{sec:coadds}

Combining WISE and NEOWISE data uniformly across all mission phases, we have built a custom set of time-domain 3-5$\mu$m coadds optimized for detecting long-timescale ($\tau \gtrsim $ 0.5 yr) motion and variability of faint sources. We stack the $\gtrsim$12 single exposures at each sky location during each sky pass, thereby obtaining one ``coadd epoch'' in each of the W1 and W2 channels every six months during which WISE has been operational. We refer to these image stacks as ``time-resolved unWISE coadds''. Each time-resolved unWISE coadd is 2048 pixels by 2048 pixels, with a pixel scale of 2.75$''$/pix, an angular extent of $1.56^{\circ}$ on a side, and a solid angle of 2.45 square degrees.

\begin{figure}
\begin{center}
\includegraphics[scale=0.575]{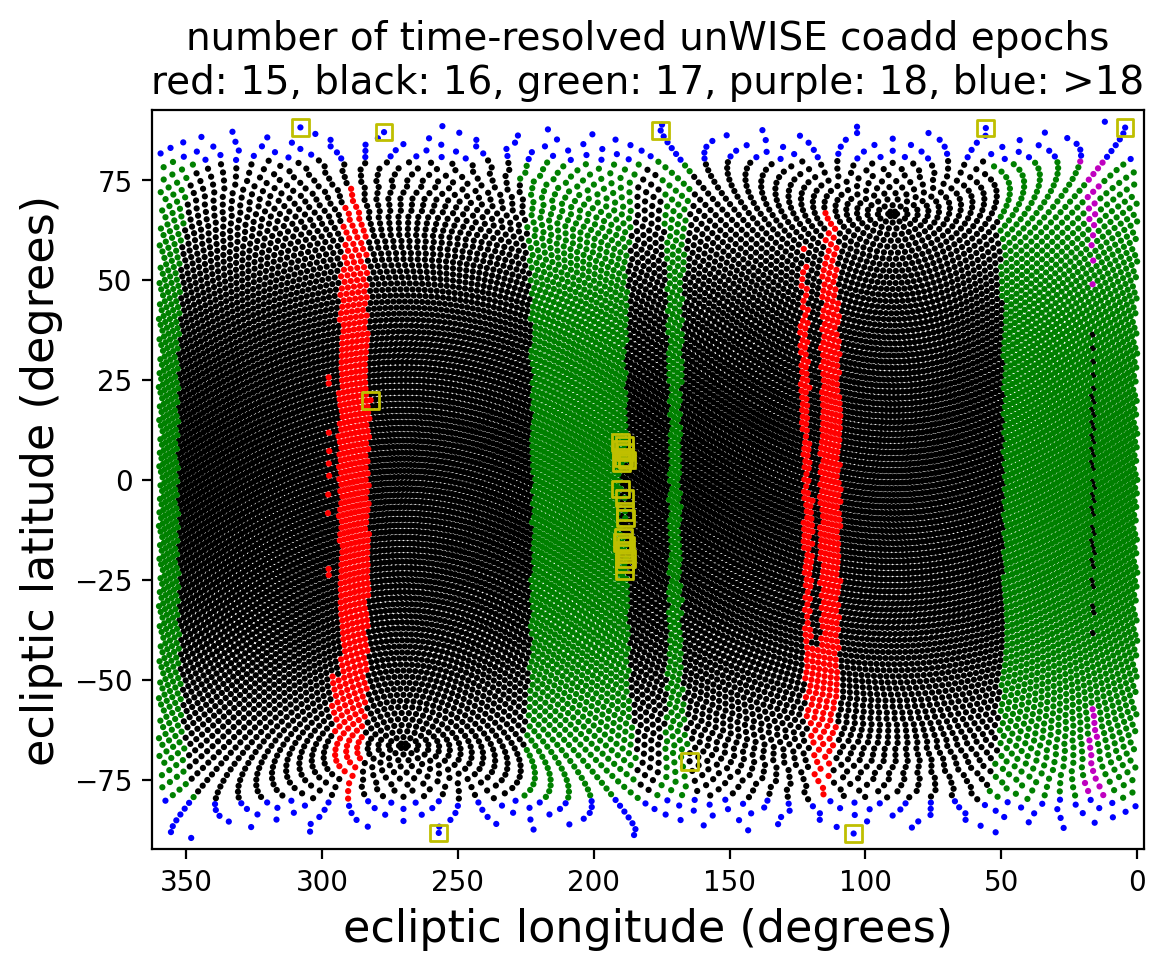}
\caption{Number of W1 coadd epochs per coadd\_id astrometric footprint (ecliptic coordinates). Typically there are 16 coadd epochs available in W1 (black dots). The red-colored ranges of ecliptic longitude have only 15 coadd epochs per band because they were impacted by the 2014 April WISE command timing anomaly. The green-colored ranges of ecliptic longitude have 17 coadd epochs in W1. The purple-colored regions have 18 coadd epochs in W1. The ecliptic poles ($|\beta| > 80^{\circ}$) are blue, indicating $>18$ available coadd epochs in W1 as a result of the modified unWISE time-slicing rules employed for these regions \citep{tr_neo2}. Yellow boxes signify coadd\_id footprints for which the number of W1 coadd epochs differs from the number of W2 coadd epochs. \label{fig:n_coadd_epochs}}
\end{center}
\end{figure}

In \cite{tr_neo3} we publicly released a full-sky set of typically 8 such coadd epochs per band per sky location, based on the first 4 years of available W1/W2 data (2010 January through 2016 mid-December). Here, we use
an updated set of time-resolved unWISE coadds that folds in an additional 4 years of NEOWISE exposures acquired between 2016 December 13 and 2020 December 13. This new full-sky set of time-resolved unWISE coadds was generated using the same code and output data model as in \cite{tr_neo3}, and now features 16 coadd
epochs per band per typical sky location (see Figure \ref{fig:n_coadd_epochs}). Most sky locations have a time baseline of 10.5 years (early-mid 2010 to mid-late 2020). The coadds include $\sim$8 years of WISE/NEOWISE observations (early 2010 to early 2011 and late 2013 through late 2020; note the hibernation period from early 2011 to late 2013 during which time WISE did not acquire science data).

At a given sky location, there are never fewer than 15 coadd epochs available per band, and the maximum number of coadd epochs per band is 295, near the ecliptic poles. The total number of single-band sets of unWISE time-resolved coadds used for this study is 616,838: 308,424 (308,414) in W1 (W2)\footnote{The number of single-band coadds can differ between W1 and W2 because unWISE coaddition performs exposure-level data quality cuts which may, on rare occasions, exclude all exposures in one band while retaining nonzero coverage in the other band.}. Therefore, we expect the unTimely Catalog to correspondingly consist of 616,838 per-band epochal catalogs. As there are 18,240 unique astrometric footprints defining the full-sky tiling used by unWISE \citep{cutri12, lang14}, this epochal coadd data set can be thought of as containing the equivalent of roughly 616,838/18,240 = 33.8 all-sky maps.

Epochal unWISE coadds and their corresponding unTimely catalogs are each identified by a unique (band, epoch, coadd\_id) triplet, where band is an integer (1 for W1, 2 for W2), epoch is a zero-indexed integer counter that increases with time, and coadd\_id is a string encoding the tile center's equatorial coordinates (e.g., coadd\_id = 1497p015 is centered at $\alpha$ = 149.748462$^{\circ}$, $\delta$ = 1.514444$^{\circ}$). In this scheme, (band, epoch, coadd\_id) = (1, 0, 1497p015) is the earliest W1 coadd epoch for tile footprint 1497p015,  (1, 1, 1497p015) is the second earliest W1 coadd epoch for tile footprint 1497p015, and so forth. More details about the unWISE time-resolved coadds can be found in \cite{tr_neo2}, which also uses this same (band, epoch, coadd\_id) notation/convention.

\section{\texttt{crowdsource} Processing} \label{sec:crowdsource}

Due to the $\sim$6.5$''$ FWHM point spread function (PSF) at W1/W2, blending is prevalent in WISE imaging across all Galactic latitudes. This ubiquitous blending strongly influences our approach for cataloging the time-resolved unWISE coadds. The \verb|crowdsource| crowded field point source photometry module \citep{decaps,crowdsource_ascl} is an excellent match to the cataloging task at hand, as it performs PSF modeling for thousands of point sources simultaneously, resulting in excellent deblending performance \citep{decaps, decaps2}. \verb|crowdsource| has previously been applied to (static sky) unWISE coadds to create the `unWISE Catalog' \citep{unwise_catalog}.

We deployed essentially the same version of \verb|crowdsource| used for the unWISE Catalog \citep{unwise_catalog} on the entire set of time-resolved unWISE coadds described in $\S$\ref{sec:coadds}. Each time-resolved coadd is processed entirely independently of all other coadds --- there is no multi-band component of the cataloging, nor is there any forced photometry. In contrast to the unWISE Catalog processing, for the unTimely Catalog we did not make use of the neural network based flagging of `nebulous' regions affected by Milky Way dust \citep{decaps,unwise_catalog}.

We inherit the unWISE Catalog's column definitions\footnote{\url{https://catalog.unwise.me/catalogs.html}} with only a few minor modifications. We also inherit three newly added columns (XISO, YISO, FLUXISO) from \verb|crowdsource| updates implemented after unWISE Catalog processing\footnote{\url{http://decaps.skymaps.info/catalogs.html}}. We updated the format of the UNWISE\_DETID column's values to encode the unWISE coadd epoch number. For example, the one thousandth detection cataloged for (band = 1, epoch = 5, coadd\_id = 1497p015) has UNWISE\_DETID = 1497p015w1o0000999e005. Also, the FLAGS\_INFO bitmask column's nebulosity bit (2$^5$) is always zero, because we did not use the \verb|crowdsource| nebulosity classifier (but see $\S$\ref{sec:dr} for the alternative nebulosity labeling mechanism employed within our unTimely Catalog data release). Lastly, we added seven new metadata columns (COADD\_ID, BAND, EPOCH, FORWARD, MJDMIN, MJDMAX, MJDMEAN), mostly related to the time-domain aspect of the unTimely Catalog. All unTimely Catalog column definitions are provided in Table \ref{tab:det}.

\begin{table*}
        \centering
        \caption{All unTimely Catalog columns.}
        \label{tab:det}
        \begin{tabular}{ll}
                \hline
                Column & Description \\
                 \hline
                X & x coordinate within unWISE coadd image (pixels) \\
                Y & y coordinate within unWISE coadd image (pixels) \\
                FLUX & PSF-fit flux in units of Vega nanomaggies; mag$_{\textrm{Vega}}$ = 22.5 $-$ 2.5$\times$log$_{10}$(FLUX) \\
                DX & uncertainty in X (pixels) \\
                DY & uncertainty in Y (pixels) \\
                DFLUX & uncertainty in flux (Vega nanomaggies; statistical only) \\
                QF & ``quality factor'' \\
                RCHI2 & average $\chi^2$ per pixel, weighted by the PSF \\
                FRACFLUX & fraction of flux in this object's PSF that comes from this object \\
                FLUXLBS & local-background-subtracted flux (Vega nanomaggies) \\
                DFLUXLBS & uncertainty in local-background-subtracted flux (Vega nanomaggies) \\
                FWHM & full-width at half-maximum of the PSF (pixels) \\
                SPREAD\_MODEL & Source Extractor \citep{source_extractor} spread\_model parameter \\
                DSPREAD\_MODEL & uncertainty in SPREAD\_MODEL \\
                FLUXISO & flux derived from linear least squares fit to neighbor-subtracted image (Vega nanomaggies) \\
                XISO & x coordinate derived from linear least squares fit to neighbor-subtracted image (pixels) \\
                YISO & y coordinate derived from linear least squares fit to neighbor-subtracted image (pixels) \\
                SKY & local sky level (Vega nanomaggies per unWISE coadd pixel) \\
                RA & right ascension (degrees); inherits L1b astrometry \\
                DEC & declination (degrees); inherits L1b astrometry \\
                COADD\_ID & unWISE coadd astrometric footprint identifier (see $\S$\ref{sec:coadds}) \\
                BAND & WISE band; 1 for W1, 2 for W2 \\
                UNWISE\_DETID & unique detection identifier (see $\S$\ref{sec:crowdsource} for details) \\
                NM & number of WISE/NEOWISE exposures contributing to the unWISE coadd at this location \\
                PRIMARY & is the center of this source in the primary region of its coadd? \\
                FLAGS\_UNWISE & unWISE coadd flags at central pixel\tablenotemark{a} \\
                FLAGS\_INFO & additional informational flags at central pixel\tablenotemark{b} \\
                EPOCH & unWISE epoch number, as defined in $\S$\ref{sec:coadds} \\
                FORWARD & boolean --- were input frames acquired pointing forward (1) or backward (0) along Earth's orbit? \\
                MJDMIN & MJD value of earliest contributing exposure \\
                MJDMAX & MJD value of latest contributing exposure \\
                MJDMEAN & mean of MJDMIN and MJDMAX \\
                \hline
        \end{tabular}
        \raggedright $^a$~\url{https://catalog.unwise.me/catalogs.html#flags_unwise}
        
        \raggedright $^b$~\url{https://catalog.unwise.me/catalogs.html#flags_info}
\end{table*}

Because one of our primary science goals is to discover extremely faint and fast-moving objects, we ran the unTimely cataloging with lower than usual detection thresholds: S/N = 4.0 (2.5) in W1 (W2). We pushed the threshold so low in W2 because we suspect that the most extreme as-yet overlooked objects in the solar neighborhood are probably quite cold \citep[e.g., Y dwarfs and late-T/Y subdwarfs;][]{cushing_y_dwarfs, kirkpatrick11, luhman_0806, burgasser_subdwarf_overview, primeval_t_dwarfs, esdTs, the_accident, meisner_t_subdwarfs, lodieu_w1810}, and hence would emit more flux at W2 than W1. Having chosen the S/N = 2.5 threshold for W2, we then selected a corresponding W1 threshold (S/N = 4.0) that yielded a source density roughly matching that obtained in W2 given its 2.5$\sigma$ threshold.

In all, 616,806 epochal unTimely catalogs were generated, 308,409 in W1 and 308,397 in W2 --- very nearly one per input time-resolved unWISE coadd (see $\S$\ref{sec:coadds}). The few cases of `missing' unTimely catalogs (15 such occurrences in W1, 17 in W2) are instances of so-called ``partial coadds''. These are cases where the WISE survey strategy and unWISE time slicing have conspired to leave a time-resolved coadd with regions of missing coverage (see \citealt{tr_neo2}, particularly their $\S$5.2, for further explanation/illustration of the partial time-resolved coadd phenomenon). The 32 missing catalog cases all correspond to nearly empty unWISE coadds, each with fewer than 1,000 pixels out of 2048$^2$ = 4.19 million having nonzero integer coverage i.e., a fractional populated area less than $2.3 \times 10^{-4}$. With $\leq 2$ square arcminutes of observations available per such time-resolved coadd, it is reasonable that no corresponding unTimely source catalogs were generated.

\begin{figure}
\begin{center}
\includegraphics[scale=0.50]{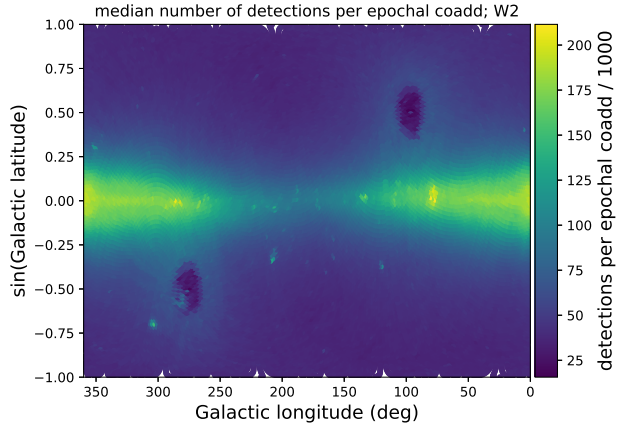}
\caption{All-sky map (Galactic coordinates) of the median number of unTimely Catalog detections per time-resolved unWISE coadd in W2. The analogous plot for W1 looks materially the same, though with number count values scaled up by $\sim$18\% overall. The Galactic plane is clearly visible. The two low density regions at ($l_{gal}, b_{gal}$) $\approx$ (96$^{\circ}$, +30$^{\circ}$) and  ($l_{gal}, b_{gal}$) $\approx$ (276$^{\circ}$, $-30^{\circ}$) are the north and south ecliptic poles, respectively. The unWISE coadd time-slicing rules change near the ecliptic poles \citep[$|\beta| > 80^{\circ}$;][]{tr_neo2}, resulting in many coadds with large regions of zero coverage, which imprints as a relatively low source density in this visualization. \label{fig:source_density}}
\end{center}
\end{figure}

Our creation of the unTimely Catalog used a total of 226,620 CPU hours\footnote{Intel Xeon ``Haswell'' processors.}: 112,823 (113,797) in W1 (W2). A map of the number of W2 sources per coadd\_id footprint is shown in Figure \ref{fig:source_density}. The total number of unTimely Catalog extracted detections is 43,459,534,445: 23,538,460,814 (19,921,073,631) in W1 (W2). The median number of unTimely Catalog detections per time-resolved unWISE coadd is 56,010 (50,574) in W1 (W2). The mean number of unTimely Catalog detections per time-resolved unWISE coadd is 76,322 (64,596) in W1 (W2). The maximum number of unTimely Catalog detections per time-resolved unWISE coadd is 272,062 (238,767) in W1 (W2). The minimum number of unTimely Catalog detections per time-resolved unWISE coadd is 2, in both W1 and W2. Such cases coincide with partial coadds that consist almost entirely of regions with no frame coverage but still manage to yield an output \verb|crowdsource| catalog file.

\section{Completeness \& Reliability} \label{sec:completeness}

\begin{figure*}
\begin{center}
\includegraphics[scale=0.40]{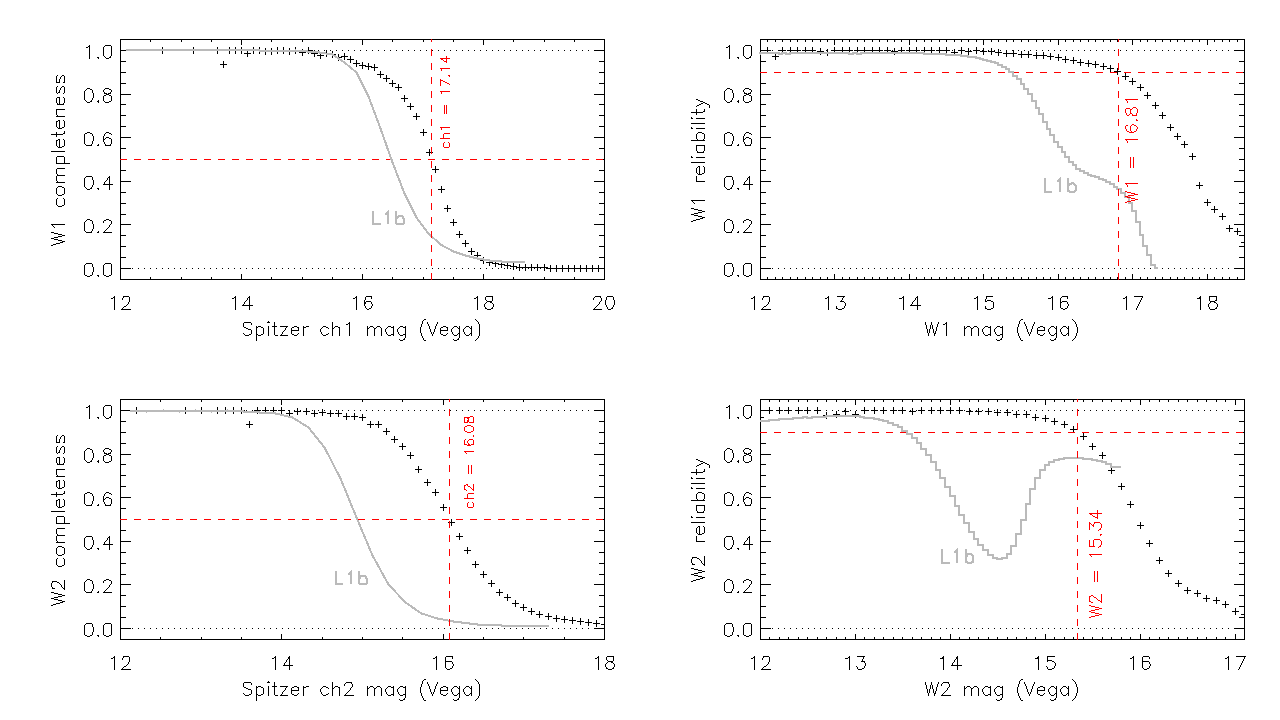}
\caption{Differential completeness and reliability results for the unTimely Catalog (black plus marks) and for the WISE/NEOWISE Single Exposure (L1b) Source Tables (gray lines). The unTimely differential completeness and reliability are assessed via comparison against the Spitzer S-COSMOS catalog \citep{scosmos}, as COSMOS is representative of the sky at low ecliptic latitude and high Galactic latitude. Top left: W1 differential completeness. Top right: W1 differential reliability. Bottom left: W2 differential completeness. Bottom right: W2 differential reliability. Red annotations provide the unTimely 50\% completeness and 90\% reliability limits in the left and right column panels, respectively. The unTimely curves shown average together the results for all 16 unTimely coadd epochs per band in the COSMOS sky region. The plotted Single Exposure Source Table values average together the per-year curves from 2010-2020, to match unTimely's temporal extent. The Single Exposure Source Table completeness values were tabulated as a function of W1 mag and W2 mag, but have been plotted in the left column as if W1 = ch1 and W2 = ch2.\label{fig:completeness}}
\end{center}
\end{figure*}

We assess the unTimely Catalog's differential completeness and reliability\footnote{By differential reliability, we mean the fraction of unTimely detections in a given narrow unTimely magnitude bin that have counterparts in a deeper ``truth'' catalog.} by comparison against deeper, higher angular resolution (FWHM $\approx 1.7''$) Spitzer imaging in the COSMOS region \footnote{\url{https://cosmos.astro.caltech.edu/}}. COSMOS lies near the ecliptic plane and at high Galactic latitude, and therefore is representative of `typical' sky locations. Our completeness/reliability analysis proceeds analogously to $\S$5.1 of \cite{unwise_catalog} and inherits the detailed parameter choices made therein, such as a 2$''$ Spitzer-WISE cross-match radius. Figure \ref{fig:completeness} shows the results of our unTimely Catalog differential completeness and reliability analyses. Figure \ref{fig:completeness} also overplots differential completeness and reliability curves for the WISE/NEOWISE Single Exposure Source Tables\footnote{\url{https://wise2.ipac.caltech.edu/docs/release/neowise/expsup/sec2_1e.html}}. All magnitudes quoted in this section are in the Vega system.

We find 50\% unTimely completeness thresholds of 17.14 mag (16.08 mag) in W1 (W2) and 90\% unTimely reliability thresholds of 16.81 mag (15.34 mag) in W1 (W2). The WISE/NEOWISE Single Exposure Source Table curves reach 50\% completeness at 16.48 mag (14.94 mag) in W1 (W2) and 90\% reliability at 15.35 mag (13.55 mag) for W1 (W2)\footnote{The WISE/NEOWISE Single Exposure Source Table reliability values quoted here and shown in Figure \ref{fig:completeness} are based on the NEOWISE Explanatory Supplement's ``moderate filtering" approach that requires \texttt{w?frtr} like `00\%'.}. Therefore, unTimely pushes 0.66 mag (1.14 mag) fainter than the Single Exposure Source Tables in terms of 50\% completeness in W1 (W2), and 1.46 mag (1.79 mag) fainter in terms of 90\% reliability. The unTimely depth enhancements compared to the Single Exposure Source Table will generally be even larger at higher eclipitic latitudes than the COSMOS values derived here, because of increasing WISE/NEOWISE frame coverage per sky pass at higher $|\beta|$ (for reference, the COSMOS field is at $|\beta|$ = 10.1$^{\circ}$). One exception to this trend is that very near the ecliptic poles ($|\beta| > 80^{\circ}$), there can often be areas of low (or even zero) coverage in particular time-resolved unWISE coadds due to the modified unWISE time-slicing rules employed in those regions.

\section{Zeropoint Consistency and \\ Bright End Scatter}
\label{sec:zp}

We investigate the unTimely Catalog's photometric zeropoint consistency from epoch to epoch and its bright end photometric scatter using a set of four unWISE tile footprints covering the COSMOS region ($\sim$10 square degrees in total). Over this region, we select a bright star sample to study for each band. We use the deep, static-sky unWISE Catalog \citep{unwise_catalog} for our bright star selections. We require FLAGS\_UNWISE = 0, FLAGS\_INFO = 0, FRACFLUX $>$ 0.99 and PRIMARY = 1 in the unWISE Catalog. We also limit to Vega magnitude ranges of [9.5, 10.5] ([8.7, 10.5]) in W1 (W2). These magnitude ranges correspond to bright but unsaturated sources and our selections yield samples of 167 (212) stars in W1 (W2).

\begin{figure*}
\gridline{\fig{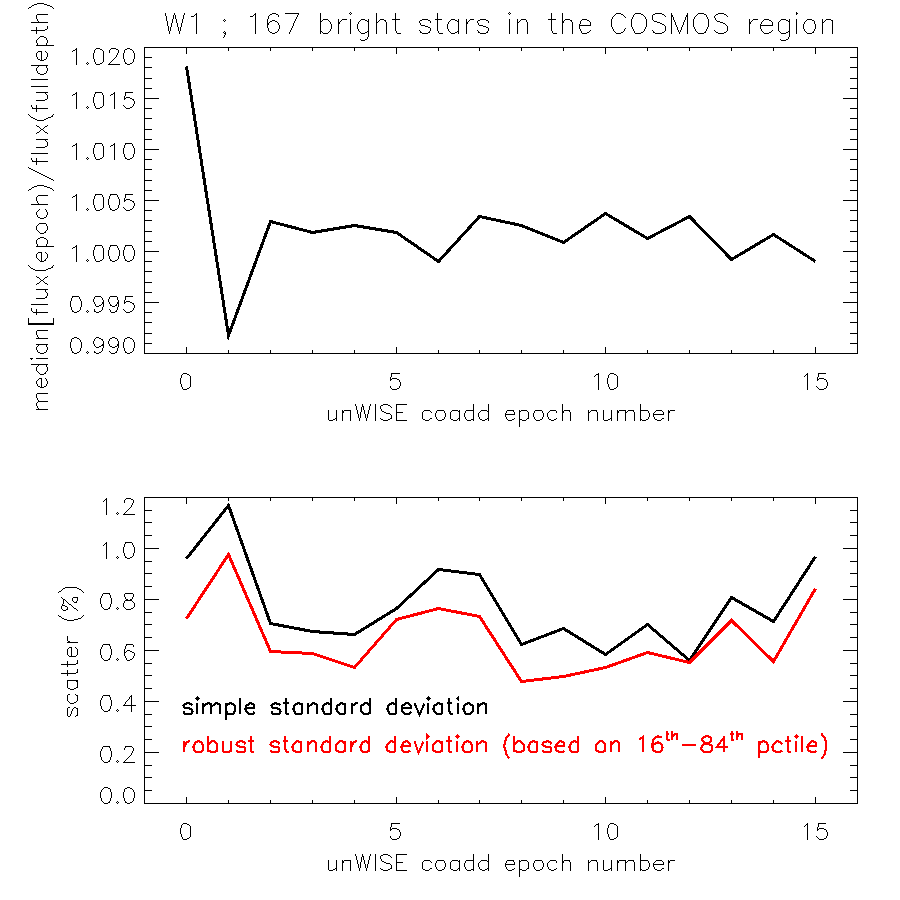}{0.50\textwidth}{(a)}
          \fig{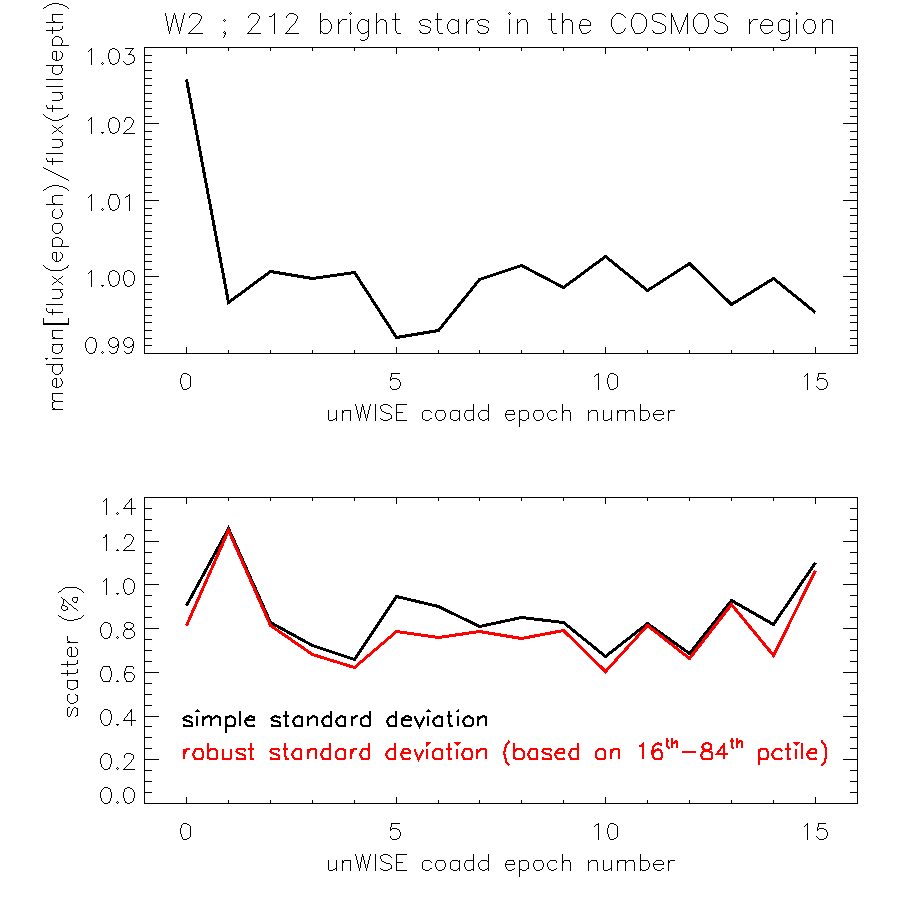}{0.50\textwidth}{(b)}
          }
\caption{Photometric zeropoint stability and bright end scatter, assessed using a sample of bright but unsaturated stars in a set of four unWISE tiles ($\sim$10 square degrees total) encompassing the COSMOS region. (a) W1 (b) W2. Top panels: in both bands, the zeropoint during the very first WISE sky pass appears to have fluxes consistently $\sim$2\% brighter than in all following sky passes, which remain consistent with one another at the sub-percent level. Bottom panels: in both bands, the bright end scatter is $\lesssim$1\%.
\label{fig:zp}}
\end{figure*}

For each selected star in each unTimely catalog epoch, we compute the ratio of its epochal flux to that of its counterpart in the static-sky unWISE Catalog. We then bin these ratios by unTimely epoch number and plot the median ratio versus epoch number in the top two panels of Figure \ref{fig:zp}. During the very first WISE sky pass, the W1 (W2) bright stars appear to have fluxes consistently $\sim$1.8\% ($\sim$2.5\%) brighter than in all following sky passes, which remain consistent with one another at the $\lesssim$1\% level. We speculate that the $\sim$2+\% unTimely zeropoint offsets seen in the first WISE sky pass may be due to imperfections in the WISE/NEOWISE zeropoint determination procedure presented in \cite{fulldepth_neo1}, particularly its assumption that zeropoints computed near the ecliptic poles can be applied without modification to the rest of the sky.

We also compute, for each epoch in each band, the scatter among the per star ratios of unTimely epochal flux to static-sky flux. These bright end scatter values are shown in the bottom panels of Figure \ref{fig:zp}, and are $\lesssim$1\%. Users with science applications that may demand photometric zeropoint consistency at the better than $\pm$2\% ($\pm$3\%) level in W1 (W2) should consider computing small zeropoint recalibration offsets on a per (band, epoch, coadd\_id) triplet basis, to remove features like the $\sim$2+\% offset seen here during the first sky pass. We have not computed an all-sky lookup table of photometric zeropoint recalibration parameters, though we may do so for future unTimely data releases.

\section{Example Science Applications} \label{sec:sci_apps}

\subsection{Faint, Fast-Moving Objects} \label{sec:moving}

\subsubsection{Discovering New Members of the Solar Neighborhood}

Of particular interest will be searching the unTimely Catalog for extremely faint and fast-moving objects in the solar neighborhood, which may be lurking very close to the Sun but have so far evaded detection due to their cold temperatures \citep[e.g.,][]{j0855}. Despite extensive surveying in recent years \citep[e.g.,][]{backyard_worlds, catwise_p14034, byw_spitzer} and full-sky astrometry from Gaia \citep{gaia_mission,gaia_dr2,gaia_dr3}, Barnard's Star \citep{barnards_star} still stands as the highest known proper motion source ($\mu \approx 10.4$ asec/yr), Proxima Centauri  remains our nearest known stellar or substellar neighbor ($d = 1.3$ pc), and WISE 0855 \citep[$T_{\textrm{eff}} \approx 250$ K;][]{j0855} still represents the coldest known brown dwarf.

It is perhaps especially surprising that no brown dwarfs as cool or cooler than WISE 0855 have thus far been discovered by recent unWISE-based searches such as Backyard Worlds and CatWISE \citep[e.g.,][]{marocco2019,j1446}. Owing to enhanced depth versus individual WISE exposures, these searches should have yielded between 4 and 35 discoveries comparable to WISE 0855 \citep[16$^{\textrm{th}}$-84$^{\textrm{th}}$ percentile, with a median of 15;][]{wright_0855}. \cite{davy_20pc_update} commented on this dearth of new WISE 0855 analogs, with one possible resolution being that deep WISE/NEOWISE catalogs like unWISE Catalog \citep{unwise_catalog} and CatWISE2020 \citep{catwise_data_paper,catwise2020} perform their source detection steps on static sky coadds spanning $\sim$5-10 year time baselines. This source detection methodology has the potential to `smear out' any signal from very faint/fast-moving objects. unTimely avoids this pitfall by performing source detection on epochal coadds, and so provides renewed hope of finding superlative moving objects, for instance cooler than WISE 0855 or with higher proper motion than Barnard's Star. During the $\sim$1 day timespan of a time-resolved unWISE coadd, even a 10$''$/yr total proper motion corresponds to a negligible `smearing' of 27 mas, equivalent to  $\sim 4 \times 10^{-3}$ PSF FWHM.

unTimely can be used to optimize unWISE-based PSF subtraction searches for faint brown dwarf companions to nearby stars, by providing a best-fit flux and centroid for the primary during each WISE sky pass. The unTimely Catalog may also be a valuable proving ground for advanced (machine learning) moving object detection linking algorithms beyond standard approaches like DBSCAN \citep{dbscan}.

\begin{figure*}
\begin{center}
\includegraphics[scale=0.375]{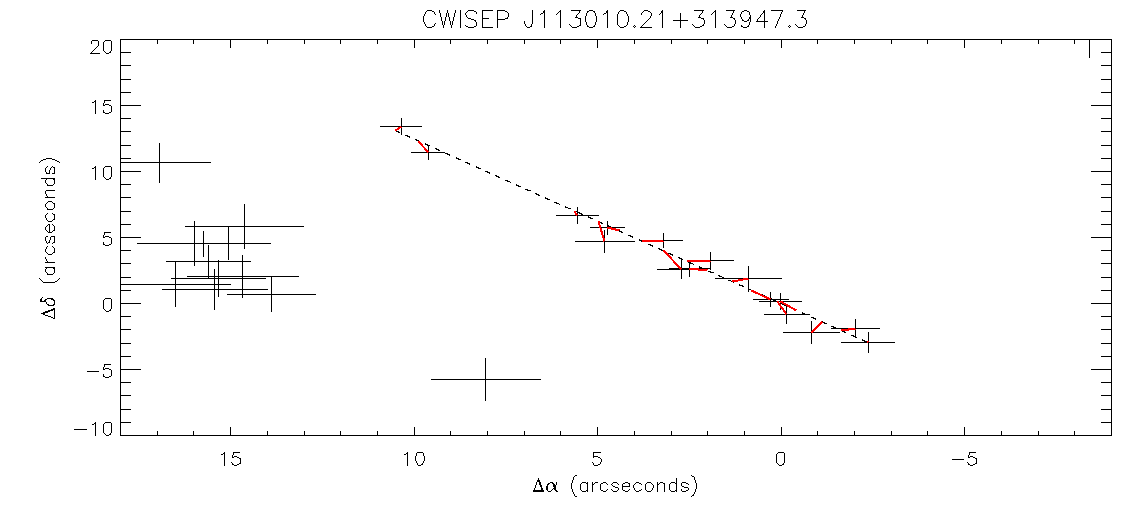}
\caption{All S/N $> 3$ unTimely Catalog W2 detections (black plus marks) in the vicinity of the faint (W2 $\approx$ 15.4 mag) and fast-moving brown dwarf CWISEP J1130+3139 \citep{catwise_p14034}. The size of each black plus mark indicates its $\pm$1$\sigma$ positional uncertainties. The dotted black line represents a linear proper motion fit based on \cite{catwise_p14034}. CWISEP J1130+3139 is moving toward the southwest (bottom right) over time. Red lines connect the 16 detections within 1 unWISE pixel (2.75$''$) of the linear motion trajectory with their corresponding epoch's location on the linear trajectory. The cluster of detections centered at $(\Delta\alpha, \Delta\delta) \approx (15'', 2.5'')$ is a faint background galaxy with W2 = 16.65 mag. The isolated detection near $(\Delta\alpha, \Delta\delta) = (8'', -6'')$ is a very low significance noise detection extracted from a single unWISE epochal coadd image. In this visualization, $(\Delta\alpha, \Delta\delta)$ = (0, 0) is at ($\alpha$, $\delta$) = (172.54118$^{\circ}$, 31.66166$^{\circ}$). \label{fig:mover}}
\end{center}
\end{figure*}

The unTimely Catalog is effectively optimized to find objects with exceptionally high WISE reduced proper motions \citep{reduced_proper_motion}, a useful indicator of extremely low luminosity in the absence of trigonometric parallax measurements. Several sought after low-temperature populations have characteristically high reduced proper motions: Y dwarfs, (extreme) T type subdwarfs \citep{esdTs} and, more speculatively, Y-type subdwarfs. Figure \ref{fig:mover} illustrates the tracklet of many unTimely Catalog detections for the faint, fast-moving T-type subdwarf candidate WISE J1130+3139 \citep{catwise_p14034}. It is critical to pinpoint the coldest, lowest luminosity brown dwarfs so that JWST \citep{jwst} can observe them spectroscopically in the mid-infrared \citep[e.g.,][]{leggett_white_paper,davy_white_paper}.

\subsubsection{Characterizing Solar Neighborhood Members}

In addition to discovery of new members of the solar neighborhood, unTimely will also be useful for further detailed characterization of nearby moving objects. Given the pervasive blending present in WISE observations, many fast-moving sources are contaminated by unrelated background objects at some epochs but are entirely free of such contamination at other epochs. These situations cause problems for CatWISE and the unWISE Catalog, which combine all WISE epochs together and cannot permit selective pruning of confused epochs on a per object basis. On the other hand, the unTimely Catalog allows for custom selection of `clean' epochs/detections for every moving object, enabling improved photometric measurements and astrometric fits. Examples of recent solar neighborhood discoveries that benefit from such treatment are WISE 0855, Ross 19b \citep{ross19b} and WISE 1810$-$1010 \citep{esdTs, lodieu_w1810}, all of which were blended with background contaminants at early WISE epochs circa 2010. unTimely Catalog can also enable a $\gtrsim$1.3 magnitude deeper version of the \cite{theissen_parallaxes} WISE parallax-fitting analysis/methodology, further filling in the set of solar neighborhood trigonometric parallaxes for cool objects not detectable with Gaia.

\subsubsection{Further Constraining Possible Saturn/Jupiter Mass Companions to the Sun}

There has been considerable past work searching for or ruling out hypothesized planets in the outer solar system at mid-infrared wavelengths \citep[e.g.,][]{luhman_planetx,p9w1,p9_3pi,iras_p9}. By grouping single-exposure WISE detections, \cite{luhman_planetx} placed the most stringent ever constraints on the presence of a Saturn/Jupiter mass companion to the Sun (``Planet X'') in the outer solar system's distant reaches. The unTimely Catalog's W2 detections can enable a $\gtrsim$1.3 magnitude deeper variant of the \cite{luhman_planetx} analysis, with a $\sim$10$\times$ longer time baseline and $\sim$8$\times$ more input W2 imaging.

\subsection{White Dwarf Science}

The unTimely Catalog provides multiple science opportunities with regard to infrared excesses around white dwarfs (WDs). Many WISE-based searches for WD infrared excesses have been performed \citep[e.g.,][]{lai_wd_excess,xu_wd_excess,dennihy17}. However, a common problem with such studies is WISE blending/confusion, which might artificially cause a mistaken infrared excess identification \citep[e.g.,][]{word_to_the_wise}. \cite{j0207} noted that such contamination can be avoided for cases of white dwarfs with significant proper motion by checking whether or not the apparent WISE excess is comoving with the WD. unTimely provides a deep and convenient data set for performing such verifications.

WD disks have also been found to display variability at WISE wavelengths \citep[e.g.,][]{xu_disk_variability}. \cite{swan19} found that most WDs with detectable dust disks show variability at the tens of percent level in multi-year WISE light curves. \cite{wd_exo_asteroid} discovered a $\sim$1 mag outburst at W1 and W2 attributed to the tidal disruption of an exo-asteroid around WD 0145+234. unTimely can push such studies of WD disk variability at 3-5$\mu$m wavelengths $\gtrsim$1.3 magnitudes fainter than would otherwise be possible with only WISE single-exposure detections.

For certain white dwarf applications involving timescales $\lesssim$~1 day \citep[e.g.,][]{hermes_wd_rotation}, the WISE/NEOWISE Single Exposure Source Tables are a more appropriate resource than unTimely, as unTimely sacrifices any information on timescales $\lesssim$~1 day for enhanced depth.

\subsection{AGN Variability} \label{sec:agn}

Active galactic nuclei (AGN) are known to vary in both the optical and infrared \citep[e.g.,][]{agn_variability_review}, though their variability has been much more thoroughly studied in the optical \citep[e.g.,][]{ps1_sdss_quasar_variability}. Large-scale population-level analyses of AGN/quasar variability remain relatively unexplored at wavelengths of 3-5~$\mu$m, with the definitive study to date performed on $\sim$10 square degrees of repeat Spitzer/IRAC imaging \citep{kozlowski}. unTimely can permit all-sky studies of long timescale AGN variability in W1/W2 to $\gtrsim$1.3 magnitudes fainter than is possible with single-exposure WISE detections, and with fewer spurious flux variations caused by artifacts like cosmic rays and satellite streaks. 

\begin{figure*}
\begin{center}
\includegraphics[scale=0.375]{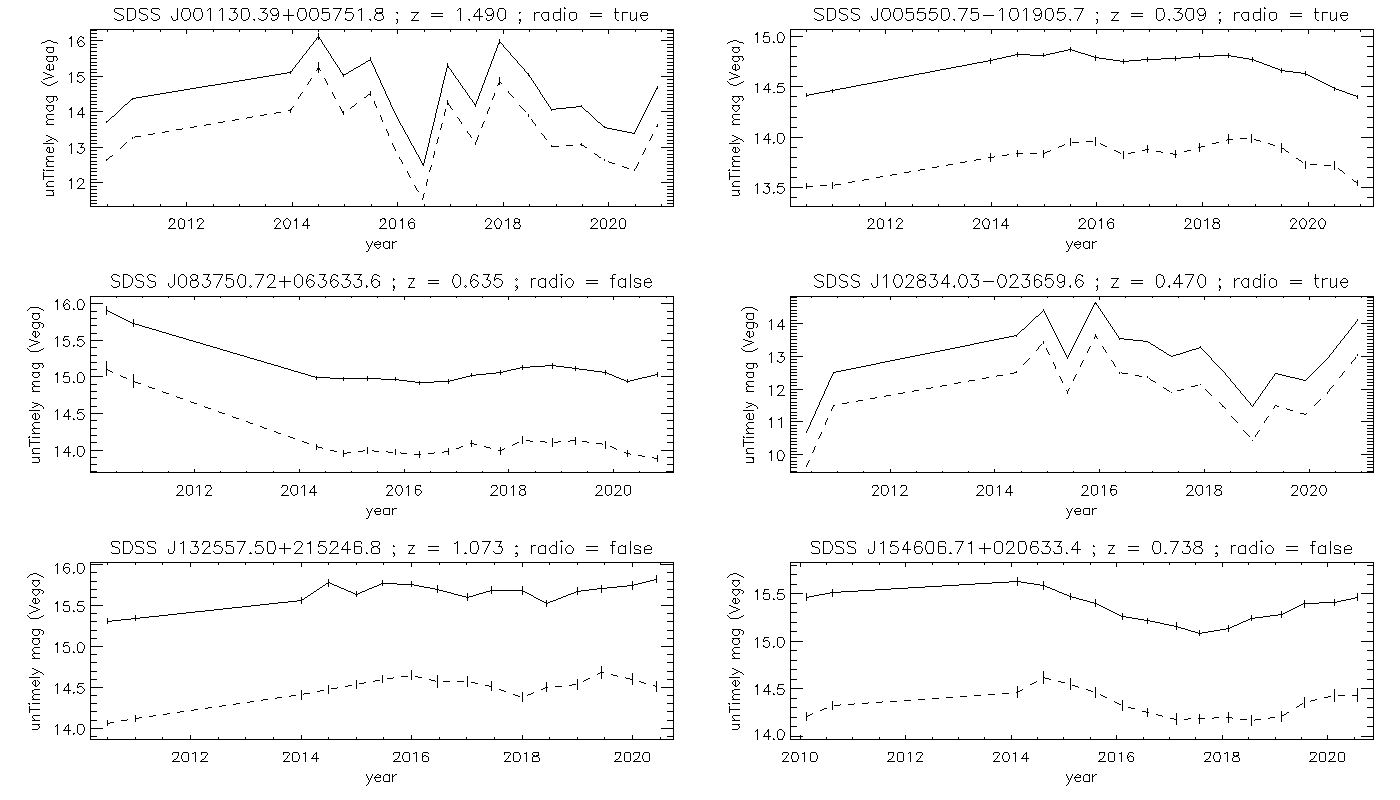}
\caption{Six infrared variable quasars drawn from the SDSS DR14 quasar catalog \citep{dr14q}. The solid (dashed) line in each panel is the unTimely Catalog W1 (W2) Vega magnitude. The ``radio = true'' or ``radio = false'' annotation is determined by whether (true) or not (false) the  SDSS DR14 quasar catalog reports a FIRST \citep{FIRST} counterpart. A lack of FIRST counterpart could indicate either a FIRST non-detection or a sky region outside the FIRST footprint. We visually inspected each of these quasars using WiseView \citep{wiseview} to validate that the variability shown is not spurious. Note that the unTimely Catalog reports only fluxes, not magnitudes. The unTimely FLUX column's units are Vega nanomaggies, which are also the native flux units for the unWISE Catalog (see \url{https://catalog.unwise.me/catalogs.html} for details). \label{fig:var_qsos}}
\end{center}
\end{figure*}

Beyond allowing detailed study of variability for tens of thousands of spectroscopically confirmed quasars, unTimely will also enable searches for rare cases of high amplitude mid-infrared AGN variability. Examples of the populations which can be thus probed include blazars and changing-look quasars \citep[e.g.,][]{wise_blazar_variability,ross2018, stern2018, qian_yang}. Figure \ref{fig:var_qsos} shows examples of large amplitude W1/W2 quasar/blazar variability in the unTimely Catalog, for a set of AGN spectroscopically confirmed by SDSS \citep{dr14q}.

\subsection{Other Flux Variables and Transients} \label{sec:variables}

Additional flux variability classes/topics for which the unTimely Catalog should prove relevant include:

\begin{itemize}
\item FU Orionis stars \citep[e.g.,][]{fu_ori_star} and young stellar object outbursts more generally \citep[e.g.,][]{lucas_yso_outburst,hillenbrand21,guo_ysos}.
\item Searching for and/or constraining mid-infrared variability of L, T and Y brown dwarfs \citep[e.g.,][Brooks et al., submitted]{bd_variability_review}.
\item Mira variables and other long period variables \citep[e.g.,][]{chen_wise_variables, mira_variables_example, wise_lpvs}.
\item Tidal disruption events and nuclear transients more broadly \citep[e.g.,][]{jiang_nuclear_transient,nuclear_transient,jiang_tde_2021}.
\item Accreting disks around young M dwarfs and/or ultracool dwarfs \citep[e.g.,][]{m_dwarf_disc, liu_ultracool_disks}.
\item Superluminous supernovae \citep[e.g.,][]{slsne_example}.
\item Searching for additional instances of ``blinking giant'' stars \citep{blinking_giant}.
\item Periodic and episodic mid-infrared brightening in Wolf-Rayet stars \citep[e.g.,][]{williams19,williams_lmc}.
\item Previously unobserved classes of infrared transients/variables \citep[e.g.,][]{spirits_overview}.

\end{itemize}

\section{Data Release} \label{sec:dr}

We release the unTimely Catalog as a set of FITS files organized within a hierarchy of directories\footnote{\url{https://portal.nersc.gov/project/cosmo/data/unwise/neo7/untimely-catalog}}. The main contents of the data release are 616,806 epochal catalog files named according to the following pattern:

\begin{center}
$<$COADD\_ID$>$\_w$<$BAND$>$\_e$<$EPOCH$>$.cat.fits.gz
\end{center}

Where, in this context, $<$COADD\_ID$>$ is an 8-character string (for instance, ``1497p015''), $<$BAND$>$ is a 1-character string version of the WISE band (either ``1'' or ``2'') and $<$EPOCH$>$ is a 3-character (left) zero-padded string encoding the epoch number (e.g., ``008'' for epoch 8, ``015'' for epoch 15, ``101 for epoch 101). As a concrete example, epoch 12 of coadd\_id = 1497p015 in W2 has a catalog basename of 1497p015\_w2\_e012.cat.fits.gz.

The catalogs for each coadd\_id astrometric footprint are co-located within a subdirectory named coadd\_id, itself within a subdirectory specified by the first 3 characters of the coadd\_id. For example, the coadd\_id = 1497p015 unTimely catalog files (for all epochs and both bands) reside within the 149/1497p015 subdirectory.

We provide an index table called untimely\_neo7\_index.fits.gz with 616,806 rows, one per cat.fits.gz catalog file. Table \ref{tab:index} provides full column definitions for this index table, which includes a column called CATALOG\_FILENAME specifying the relative path of each catalog file within the unTimely data release.

As mentioned in $\S$\ref{sec:crowdsource}, we ran \verb|crowdsource| without employing its nebulosity flagging capabilities. Instead, we have incorporated the deeper static sky nebulosity masking information from the 6-year version of the unWISE Catalog\footnote{\url{https://portal.nersc.gov/project/cosmo/data/unwise/neo5/unwise-catalog/iminfo}} into the unTimely index table. The unWISE Catalog's nebulosity masking breaks each of the 18,240 coadd\_id footprints into 64 non-overlapping 256 pix $\times$ 256 pix patches and labels each of these as being either affected ($\mathcal{N}_{i,j} = 1$) or unaffected by nebulosity ($\mathcal{N}_{i,j} = 0$). The nebulosity flagging for a given unWISE coadd\_id can therefore be encoded as an unsigned 64 bit integer in the following way:

\begin{equation}
\textrm{NEBULOSITY\_BITMASK} = \sum\limits_{i=0}^{7}\sum\limits_{j=0}^{7}\mathcal{N}_{i,j} \times 2^{8i+j}
\end{equation}

Where $\mathcal{N}_{i,j} \in [0,1]$ is the native unWISE Catalog's nebulosity mask (originally 2048 pix $\times$ 2048 pix) rebinned to an 8 $\times$ 8 image. This NEBULOSITY\_BITMASK integer value is then reported in a correspondingly named column of untimely\_neo7\_index.fits.gz, based on each row's coadd\_id. The index file also includes a convenience column called HAS\_ANY\_NEBULOSITY, which is 0 when NEBULOSITY\_BITMASK is 0 and 1 otherwise. Note that some FITS readers (such as IDL/mrdfits) may have difficulty properly parsing 64-bit unsigned integers. 9.5\% of all unTimely epochal catalogs correspond to a coadd\_id with at least one of its 64 sectors affected by nebulosity. Given that \verb|crowdsource| attempts to aggressively decompose nebulosity into a sum of point sources, which are likely deblended differently from one epoch to the next, it may be advisable to avoid catalogs with HAS\_ANY\_NEBULOSITY = 1 when performing rare object searches or all-sky population studies.

We also include the W1/W2 PSF models used by \verb|crowdsource| during unTimely processing, in a directory named \verb|psfs|. These PSF models are organized into 359 compressed tar files, one per three-digit RA subdirectory of unTimely catalog outputs.

The total unTimely Catalog data volume is 4.3 TB. Whereas the unWISE Catalog data release includes metadata image files (\url{https://catalog.unwise.me/images.html}), we exclude these from the unTimely Catalog data release to minimize the total data volume.

\begin{table*}
        \centering
        \caption{unTimely Catalog index table column descriptions.}
        \label{tab:index}
        \begin{tabular}{ll}
                \hline
                Column & Description \\
                 \hline
                BAND & integer WISE band; either 1 or 2  \\
                COADD\_ID & coadd\_id astrometric footprint identifier as defined in $\S$\ref{sec:coadds} \\
                EPOCH & \verb|epoch| number, as defined in $\S$\ref{sec:coadds} \\
                CATALOG\_FILENAME & relative path of unTimely catalog FITS file \\
                N\_DET & number of unTimely Catalog detections \\
                N\_DET\_5SIGMA & number of $\ge$5$\sigma$ unTimely Catalog detections \\
                COVMIN & minimum integer coverage in unWISE \verb|-n-u| coverage map \\
                COVMAX & maximum integer coverage in unWISE \verb|-n-u| coverage map \\
                COVMED & median integer coverage in unWISE \verb|-n-u| coverage map \\
                NPIX\_COV0 & number of pixels in \verb|-n-u| map with integer coverage of 0 frames \\
                NPIX\_COV1 & number of pixels in \verb|-n-u| map with integer coverage of 1 frame \\
                NPIX\_COV2 & number of pixels in \verb|-n-u| map with integer coverage of 2 frames \\
                FRAC\_COV0 & fraction of pixels in \verb|-n-u| map with integer coverage of 0 frames \\
                FRAC\_COV1 & fraction of pixels in \verb|-n-u| map with integer coverage of 1 frame \\
                FRAC\_COV2 & fraction of pixels in \verb|-n-u| map with integer coverage of 2 frames \\
                N\_EXP & number of exposures contributing to the coadd \\
                RA & tile center right ascension (degrees) \\
                DEC & tile center declination (degrees) \\
                LGAL & Galactic longitude corresponding to the tile center (degrees) \\
                BGAL & Galactic latitude corresponding to the tile center (degrees) \\
                LAMBDA & ecliptic longitude corresponding to the tile center (degrees) \\
                BETA & ecliptic latitude corresponding to the tile center (degrees) \\
                MJDMIN & MJD value of earliest contributing exposure \\
                MJDMAX & MJD value of latest contributing exposure \\
                MJDMEAN & mean of MJDMIN and MJDMAX \\
                DT & difference of MJDMAX and MJDMIN (days) \\
                FORWARD & boolean --- were input frames acquired pointing forward (1) or backward (0) along Earth's orbit? \\
                NEBULOSITY\_BITMASK & integer encoding which image sectors have nebulosity; see $\S$\ref{sec:dr} for details \\
                HAS\_ANY\_NEBULOSITY & 0 if NEBULOSITY\_BITMASK is 0; 1 if NEBULOSITY\_BITMASK $>$ 0 \\
                \hline
        \end{tabular}
\end{table*}

\subsection{unTimely Catalog Explorer Tool}
\label{sec:explorer}
Although the unTimely Catalog is not, as of this writing, hosted by any archive/database service, we have  built Python tools for conveniently querying and visualizing the unTimely Catalog, abstracting away the need for end users to consider details like catalog file names. We refer to this set of tools as unTimely Catalog Explorer\footnote{\url{https://github.com/fkiwy/unTimely_Catalog_explorer}}. unTimely Catalog Explorer offers capabilities to: box search query unTimely by (RA, Dec) coordinates, overlay these unTimely query results on unWISE coadd cutouts in the form of a finder chart, produce light curve check plots of the detections retrieved, and generate unWISE time series image blinks with unTimely detections overplotted. Instructions for using these unTimely Catalog Explorer functionalities are provided in the repository's \texttt{README.md} file\footnote{\url{https://github.com/fkiwy/unTimely_Catalog_explorer/blob/main/README.md}}. We note that unTimely Catalog Explorer is an excellent resource for users interested in detailed analysis of a modest number of objects, but is still not a substitute for a database service in the limit of large samples/queries.

\section{Cautionary Notes \& Potential Future Improvements} \label{sec:caveats}

Here we provide a list of cautionary notes which we expect to be relevant for unTimely Catalog end users:

\begin{itemize}

\item As mentioned in $\S$\ref{sec:crowdsource}, some time-resolved unWISE coadds have regions of very low or zero frame coverage. The unTimely photometry/astrometry in regions of low (but nonzero coverage) may be problematic. Users can flag regions of low coverage based on the NM integer coverage column provided by \verb|crowdsource|. The amount of zero (or very low) frame coverage in the time-resolved coadd corresponding to each unTimely catalog file can be assessed using the untimely\_neo7\_index.fits.gz columns N\_PIX\_COV[0-2] or FRAC\_PIX\_COV[0-2].

\item At present, there is no positional grouping of detections into celestial `objects', either across epochal catalogs or across bands.
\item unTimely Catalog deblending can be different from one epoch to another, potentially leading to spurious variability signatures in light curves. Users are advised to check for these situations upon encountering any seemingly remarkable instances of large photometric variability.
\item In terms of visually inspecting the time-resolved unWISE coadds for potential artifacts or blending, we strongly recommend using the WiseView browser-based image blinking tool \citep[\url{http://byw.tools/wiseview};][]{wiseview}.
\item Users may wish to restrict to relatively secure detections (for instance, S/N $>$ 5$\sigma$). This is a reasonable data analysis approach, but note that the simultaneous deblending of high and low significance sources means that the rejected low-significance sources may still have influenced the remaining high significance sources, potentially in different ways at different epochs.
\item Because unTimely performs source detection independently on each time-resolved coadd, light curves of objects with fluxes near the detection threshold are subject to a selection bias whereby detections are more likely to be made when the source is brighter and/or random image noise happens to enhance rather than detract from the true flux.
\item The unTimely Catalog does not incorporate W1/W2 observations from the 2022 March NEOWISE release, which made public the calendar 2021 year of NEOWISE imaging.
\item To conserve disk space, we have removed the \verb|crowdsource| ``mod" and ``info" metadata images. Ideally these could be retained, as they have been for the unWISE Catalog data release \citep{unwise_catalog}. The image-level artifact/quality bitmasks contained in the omitted unTimely ``info'' images can nevertheless be obtained from the \verb|-msk| files available in other unWISE data releases\footnote{For instance \url{https://portal.nersc.gov/project/cosmo/data/unwise/neo7/unwise-coadds/fulldepth}.}.
\item \verb|crowdsource| models images as sums of point sources --- it provides no extended galaxy profile modeling capabilities. Well-resolved galaxies are likely deblended into large numbers of point sources in ways that are not consistent from one unWISE coadd epoch to another. This may be particularly relevant for any transients embedded within large, well-resolved galaxies. This issue is at least partially mitigated thanks to \verb|crowdsource|'s implementation of less aggressive deblending within HyperLeda galaxies \citep{hyperleda, unwise_catalog}. Also, the 2$^1$ bit of unTimely's FLAGS\_INFO column labels sources that overlap with large ($d_{25} \gtrsim 7''$) HyperLeda galaxies.
\item While the time-resolved unWISE coadds largely filter out cosmic ray strikes, some residual cosmic ray imprints occasionally leak through. These residual cosmic ray imprints can appear as unTimely Catalog detections. We have found that, for S/N $>$ 10 sources in relatively uncrowded fields,  SPREAD\_MODEL $< -0.021$ identifies residual cosmic ray imprints reasonably well.
\item We have not attempted to recalibrate the unTimely Catalog astrometry to Gaia. The unTimely Catalog astrometry therefore inherits all of the astrometric imperfections/caveats discussed extensively in \cite{tr_neo2}. These astrometric systematics can reach levels of up to $\sim$250 mas per coordinate. For any applications such as parallax fitting that require high fidelity astrometry, users should perform their own astrometric calibration tweaks/checks relative to Gaia.
\item It may be possible to refine the unTimely Catalog photometry at the $\le 3$\% level by computing small zeropoint corrections on a per time-resolved coadd basis, but we have not done so as part of this data release (see $\S$\ref{sec:zp} and Figure \ref{fig:zp}).
\item At present, the unTimely Catalog only exists as a (large) set of FITS files. It would be helpful to host the unTimely Catalog in a database at an archiving facility such as NOIRLab's Astro Data Lab \citep{data_lab_aspc, data_lab_spie} and/or IRSA. It would also be useful to have an unTimely Catalog `light curve service' such that one could easily collect the light curve of a particular (stationary) object of interest --- this application would benefit from an unTimely table that groups epochal detections into objects.
\end{itemize}

\section{Conclusion} \label{sec:conclusion}

We have generated and publicly released a full-sky, time-domain unWISE catalog which we refer to as the unTimely Catalog. We expect that this catalog will be used to discover faint, fast-moving objects and probe long timescale mid-infrared variability to depths not previously attainable. We hope to address many of the opportunities for improvement outlined in $\S$\ref{sec:caveats} with future data releases of the unTimely Catalog. It would also be of interest to build a complementary version of the unTimely Catalog that operates in a pure forced photometry mode, rather than performing source detection separately on all time-resolved unWISE coadds.

\section*{Acknowledgments}

We thank the anonymous referee for helpful comments which improved this manuscript. We thank Sarah Casewell, Erik Dennihy, Siyi Xu, and Laura Rogers for input regarding potential white dwarf science applications. We thank Dustin Lang for suggesting ``unTimely'' as the name for our catalog. We thank Phil Lucas for valuable comments on a preliminary version of this manuscript. We thank Roc Cutri for sharing WISE/NEOWISE single-exposure completeness and reliability data.

\vspace{5mm}
\facilities{IRSA \citep{neowiser_frame_metadata,neowiser_l1b_source_table, irsa_atlas_images,neowiser_l1b}, 
            NEOWISE,
            NERSC,
            WISE}

\software{Astropy \citep{astropy:2013, astropy:2018},
          \texttt{crowdsource} \citep{decaps,crowdsource_ascl},
          \texttt{fkiwy/Finder\_charts} \citep{fkiwy_finders_ascl},
          IDLUTILS,
          \texttt{unwise\_psf} \citep{unwise_psf},
          WiseView \citep{wiseview}
          }

\bibliography{sample631}{}
\bibliographystyle{aasjournal}

\end{document}